\documentclass[aps,prl,twocolumn,showpacs,superscriptaddress,preprintnumbers,amsmath,amssymb]{revtex4}

\usepackage{graphics}
\usepackage{graphicx}
\usepackage{dcolumn}
\usepackage{bm}
\usepackage{amsmath}
\usepackage{rotating}
\usepackage[bookmarks = true, colorlinks = true, urlcolor = blue,hyperfootnotes=false]{hyperref}

\newcommand{\bra}[1]{\langle #1|}
\newcommand{\ket}[1]{|#1\rangle}

\renewcommand{\v}[1]{\mathbf{#1}}

\begin{document}
\title{Controlling nuclear spin exchange via optical Feshbach resonances in ${}^{171}$Yb}
\author{Iris Reichenbach}
\email{irappert@unm.edu}
\affiliation{{Department of Physics and Astronomy, University of New Mexico, Albuquerque, New Mexico 87131, USA}}
\author{Paul S. Julienne}
\affiliation{Joint Quantum Institute, NIST and University of Maryland, Gaithersburg, MD 20899-8423 USA}
\author{Ivan H. Deutsch}
\affiliation{{Department of Physics and Astronomy, University of New Mexico, Albuquerque, New Mexico 87131, USA}}

\begin{abstract}
Nuclear spin exchange occurs in ultracold collisions of fermionic alkaline-earth-like atoms due to a difference between s- and p-wave phase shifts.  We study the use of an optical Feshbach resonance, excited on the ${}^1S_0 \rightarrow {}^3P_1$ intercombination line of ${}^{171}$Yb, to affect a large modification of the s-wave scattering phase shift, and thereby optically mediate nuclear exchange forces.  We perform a full multichannel calculation of the photoassociation resonances and wave functions and from these calculate the real and imaginary parts of the scattering length.  As a figure of merit of this interaction, we estimate the fidelity to implement a $\sqrt{SWAP}$ entangling quantum logic gate for two atoms trapped in the same well of an optical lattice.  For moderate parameters one can achieve a gate fidelity of   $\sim95\% $ in a time of $\sim 50 \mu$s.
\end{abstract}

\maketitle

Interest in the use of ultracold alkaline-earth-like atoms for applications in quantum information processing has grown in recent years due to some unique features of their electronic structure and experimental advances in quantum control for optical-clock technology \cite{boyd:2006}.  Because the ground state is a closed shell ${}^1S_0$ state, for fermionic species, quantum coherence can be stored in the nuclear spins that are isolated from hyperfine coupling and very weakly coupled to perturbing noisy magnetic fields.  The nuclear spins, nonetheless, can mediate strong electronic interactions between atoms due the quantum statistics of identical particles \cite{hayes:2007}.  In addition, the existence of meta-stable ${}^3P$ excited states provides further avenues for control by transferring quantum coherence between nuclear and electronic degrees of freedom.  These various tools have been assembled in a number of proposals for quantum control, including cooling of atomic vibration without decohering nuclear spin coherence \cite{reichenbach:2007}, architectures for quantum computers \cite{daley:2009}, and quantum simulations of exotic forms of quantum magnetism \cite{gorshkov:2009}.

An important additional tool for quantum manipulations is the capability to control the interaction strength between atoms through Feshbach resonances \cite{chin:2009}. This has been essential in the study of quantum many-body phenomena in dilute ultracold gases \cite{bloch:2008, giorgini:2008}, enabling tuneable interactions for applications including the study of quantum phase transitions such as the BEC-BCS cross-over \cite{bourdel:2004}, and the production of ultracold molecules \cite{regal:2003}.  In systems of cold alkali atoms, Feshbach resonances occur in the ground electronic manifold between open and closed hyperfine channels, and are typically tuned with magnetic fields, though optical tuning has also recently been demonstrated \cite{bauer:2009}.  The lack of hyperfine structure in the ground state of the alkaline-earths removes this possibility, but the existence of the meta-stable ${}^3P$ excited states makes these atoms attractive for implementation of optical Feshbach resonances (OFR) \cite{bohn:1997, ciurylo:2005, jones:2006}, induced through laser coupling of the scattering atoms to an electronic-excited-state bound molecule.  In contrast to alkali atoms  where observation of an OFR is accompanied by substantial losses due to rapid spontaneous emission \cite{fatemi:2000, theis:2004, thalhammer:2005}, the narrow intercombination lines are ideal for use in an OFR.  Optical effects in collisions of alkaline-earth-like atoms have been demonstrated in recent experiments, including photoassociation spectroscopy in ${}^{171}$Yb \cite{enomoto:2008} and ${}^{88}$Sr \cite{zelevinsky:2006} and OFR of several bosonic isotopes  with zero nuclear spin, including ${}^{172}$Yb, ${}^{176}$Yb \cite{enomoto:2008_2} and  ${}^{88}$Sr \cite{killian:2009}. 

\begin{figure}
\begin{center}
\includegraphics[width= 0.7\columnwidth]{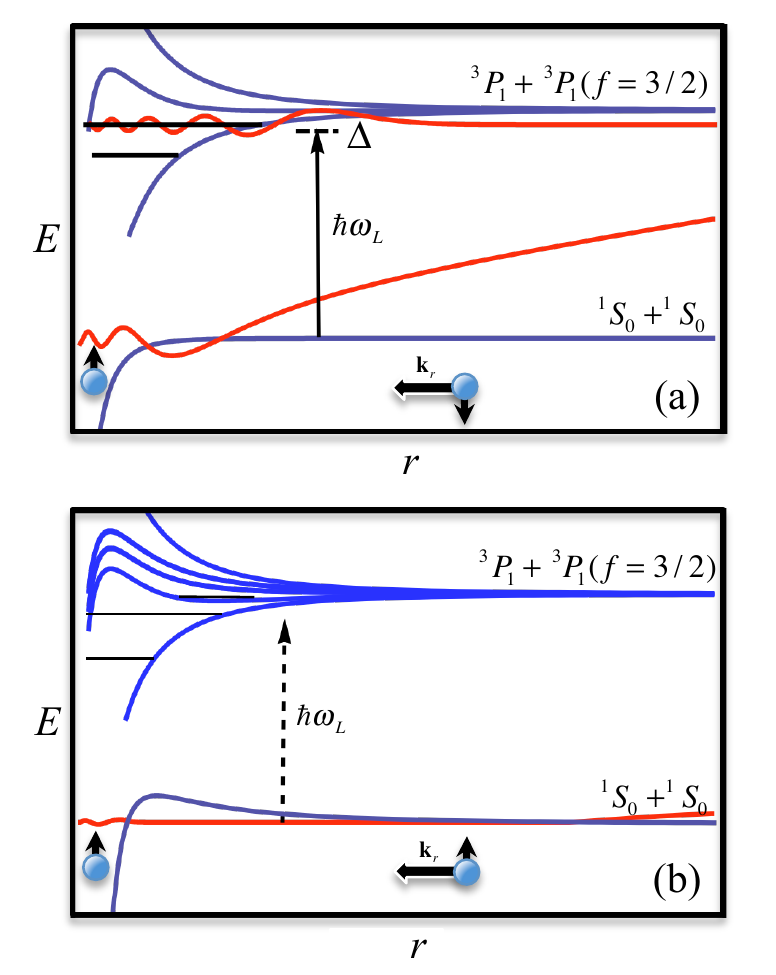}
\caption{(Color online) Schematic of the OFR, leading to nuclear spin exchange. Optical excitation of atoms colliding in an s-wave is possible only for the nuclear spin singlet (a). Nuclear spin triplet p-wave collisions (b) are suppressed at low energies and because of selection rules, since they connect to different molecular bound states (see text).}
\label{fig:schematic}
\end{center}
\end{figure}
In this letter we study OFRs  as a tool for optically controlling entangling interactions between nuclear spins of ${}^{171}$Yb atoms, a fermionic group-II-like element with a spin $i=1/2$ nucleus. This control can be accomplished through the laser modification of s-wave scattering interaction.  Because of the exchange symmetry of identical fermions, s-wave collisions (dominant at ultracold temperatures) are allowed only for the anti-symmetric nuclear-spin singlet pairing.  A relative phase between nuclear-spin singlet and triplet of $\pi/2$ arising in the collision generates an entangling $\sqrt{SWAP}$, studied previously by Hayes {\em et al.} \cite{hayes:2007}, and observed by Anderlini {\em et al.} \cite{anderlini:2007} in an equivalent scheme for hyperfine states in alkali atoms.  In this paper we show how the state of the nuclear spins allows or forbids the optical coupling of the colliding atoms to an excited molecular bound state, and thereby mediates the presence of an OFR, which strongly affects the exchange force (see Fig. 1).

We consider OFRs based on laser excitation near the strongest of the intercombination lines, ${}^1S_0 \rightarrow {}^3P_1$.  To model the Feshbach resonance, we require a good understanding of both the ground-state scattering wave function and the bound-excited-molecular wave functions. For alkaline-earth-like atoms, there is a single electronic ${}^1S_0+{}^1S_0$ ground state channel. For scattering states at low energies, the bulk of the probability amplitude of the wave functions lies away from the chemical binding region, and we need not model the ${}^1S_0+{}^1S_0$ potential in detail. Instead, we make use of the analytic solutions to a van der Waals $1/r^6$ potential found by Gao \cite{gao:1998}.  A superposition of the regular and irregular solutions, matched by the boundary conditions to achieve the correct asymptotic solution with the appropriate known scattering length \cite{kitagawa:2008}, yields an excellent approximation to the exact wave function for separations larger than $\sim 10$\AA. 

The excited states are much more complicated, giving rise to a multitude of different coupled potentials, and thus requiring a full multichannel description. The Hamiltonian for the diatomic molecule that asymptotes to unbound ${}^1S_0+{}^3P_1$ atoms has the form
\begin{equation}
\label{Hamiltonian}
 H=\frac{p_r^2}{2 \mu}+H_{rot}+H_{BO}+H_{HF}.
\end{equation}
The first two terms represent the radial and angular kinetic energy of the nuclei. $H_{BO}$ represents the electronic Born-Oppenheimer potentials that asymptote to the ${}^1S_0 + {}^3P_1$ scattering channel, taken in ``Hund's-case (c)" where the spin-orbit interaction is the dominant effect.  This is modeled through a Lennard-Jones plus dipole-dipole potential,
\begin{equation}
 \label{HBO}
 V_{\Omega\sigma}(r)=\frac{C_{12}}{r^{12}}-\frac{C_6}{r^6}-\sigma\frac{C_3^\Omega}{r^3},
\end{equation}
with parameters determined by fits to experiments as $C_6=2810$ au, $C_{12}=1.862\times10^8$ au and  $C_{3}^{1}=-C_{3}^{0}/2=0.09695$ au. $H_{HF}$ is the hyperfine interaction in the ${}^3P_1$ orbital.  We neglect the small magnetic dipole-dipole interactions between nuclei.  We do not include an external magnetic field in this calculation.

To determine the eigenstates, we perform a multichannel diagonalization. Note that for the given Hamiltonian there are only a few exact good quantum numbers:  the total angular momentum quantum number $T$, its projection on the space-fixed quantization axis $M_T$, and parity due to inversion of all particles $p$.  Other quantum numbers are approximate, depending on the dominant forces.  Three different channel bases defined by different couplings of the angular momenta of the atoms are useful in analyzing the eigenstates \cite{tiesinga:2005}.  The $H_{BO}$ is diagonal in a Hund's case (c) basis, extended to include nuclear spin, $\ket{\gamma}=\ket{J\Omega I\iota\Phi (T,M_T,p)}$, where $J,I,F$ are the magnitudes of the electron, nuclear, and total (excluding rotation) angular momentum, with projections on the internuclear axis $\Omega,\iota,\Phi$, respectively.  In contrast, the rotational and hyperfine Hamiltonians are diagonal in an extended Hunds case-(e) basis which is separable between nuclear rotation and other degrees of freedom, $\ket{\epsilon}=\ket{f_1f_2FR(T,M_T,p)}$.  Here $f_1=1/2$ and $f_2=1/2,3/2$ are the possible individual atomic angular momenta for the ground and excited state, respectively.  A third basis of importance is the product basis of internal atomic states and nuclear rotations, $ \ket{\pi}=\ket{f_1,m_1,f_2,m_2,R(T,M_T,p)}$.

Selection rules divide the excited-state channels into two classes: those that are optically accessible by ground-states colliding in an s-wave  from those accessible in a p-wave collision.  Because of the correlation between nuclear spin pairing and collisional partial wave, these selection rules play an essential role in optical control of nuclear exchange forces.  The dipole-allowed transition requires that the ground and excited states have opposite parity and that total angular momentum changes by $\Delta T =0,1$ ($T=0\rightarrow 0$ is forbidden).    By exchange symmetry, in the ground state, the two-atom nuclear spin singlet ($I=0$) is associated with s-wave collisions ($R=0$) and therefore has even parity $p=0$ and total angular momentum $T=0$. Thus the excited state must have $p=-1$ and $T=1$. Taking the selection rules into account, there are 5 excited channels, each 3-fold degenerate in the absence of an external magnetic field, which can be optically excited for an s-wave collision in the ground state.  In contrast, the two-atom nuclear spin triplet ($I=1$) is associated with p-wave collisions ($R=1$) and therefore we have odd parity and possible total angular momenta $T=0,1,2$ in the ground state.  By the selections rules, the electric-dipole connected excited states must have $T=0,1,2,3$, and even parity.  In total, the p-wave collisions are optically coupled 19, (2T+1)-fold degenerate channels in the absence of a magnetic field (a total of 89 channels).

A full multichannel diagonalization of the Hamiltonian based on the DVR method \cite{colbert:1982} yields the bound states representing photoassociation resonances that dissociate to  ${}^1S_0 + {}^3P_1 (f=3/2)$.  The s-wave-accessible resonances are shown Table  \ref{tab:swaves}. These are in good agreement with the experimental observations of Enomoto {\em et al.} \cite{enomoto:2008_2}, though some states were not reported in that experiment, either because the binding energy is comparable to the linewidth or because these state are too tightly bound and thus have small Franck-Condon factors. We also note the existence of potential curves that support so-called purely long-range (PLR) states, observed by Enomoto {\em et al.} for p-waves.  These potentials arise from avoided crossings due to hyperfine mixing of different Hund's case-(c) potentials.   Because of the weakness of this interaction, the avoided crossings occur at very large internuclear separations and the resulting potential is extremely shallow (a 59.2 MHz deep PLR potential accessible through laser excitation of an s-wave collision and a 717 MHz deep PLR potential for p-waves).  Nonetheless, they can support bound molecular states.

With the so determined wave functions and eigenenergies, we calculate the properties of our OFR. The optical modification of the s-wave  scattering length can be expressed as $a(I,\Delta)=a_{bg}+a_{opt}(I,\Delta)-ib_{opt}(I,\Delta)$ \cite{bohn:1997}, where $a_{bg}$  is the background scattering length,
\begin{equation}
 a_{opt}(I,\Delta)=l_{opt}(I)\left(\frac{\Delta\Gamma_M}{\Delta^2+(\Gamma_M+\Gamma_{stim}(I))^2/4}\right) 
 \end{equation}
is the optical contribution to the elastic scattering length, depending on the intensity $I$ of the radiation and the detuning $\Delta$, and 
\begin{equation}
 b_{opt}(I,\Delta)=\frac{l_{opt}(I)}{2}\left(\frac{\Gamma_M^2}{\Delta^2+(\Gamma_M+\Gamma_{stim}(I))^2/4}\right),
\end{equation}
is the imaginary part of the scattering length, accounting for two-body scattering loss. The molecular spontaneous decay rate, $\Gamma_M$, is calculated from the multichannel wave function.  The linewidth is broadened by stimulated emission at the rate \cite{bohn:1997},
\begin{equation}
\label{gammastim}
 \Gamma_{stim}(I)=\frac{\pi}{2}\left(\frac{I}{I_{sat}}\right)\hbar\Gamma_A^2 f_{rot}f_{FC},
\end{equation}
Here $\Gamma_A/2\pi=182$ kHz is the atomic natural linewidth for the ${}^1S_0\rightarrow{}^3P_1$ intercombination transition in ${}^{171}$Yb, $I_{sat}=(2\pi^2\hbar\Gamma_Ac)/(3\lambda^3)=0.13$ mW/cm$^2$  is the atomic saturation intensity. The Franck-Condon factor that measures the overlap between the spatial wave function of the (energy normalized) scattering ground electron state at energy $E$ and the (unit normalized) bound excited state with the appropriate rotational modification \cite{ciurylo:2005} is given by
\begin{equation} 
\label{fc}
f_{FC}f_{rot} =\left| \bra{\psi_e}\v{d}\cdot \epsilon_L \ket{\psi_g} \right|^2/2d_A^2,
\end{equation}
where $d_A$ is the atomic dipole moment for this transition and $\epsilon_L$ is the laser polarization.  Equation (\ref{fc}) is calculated using the multichannel wavefunctions expanded in the basis $\ket{\pi}$, as discussed above and in \cite{tiesinga:2005}. 

The key parameter that determines the strength of the resonance is the ``optical length" defined by $l_{opt}(I)=\Gamma_{stim}(I)/(2k_r \Gamma_M)$, where $k_r$ is the wave number for the relative motion of the reduced mass $\mu$.  Table I gives the optical length for the s-wave-accessible photoassociation resonances near dissociation, calculated at a laser intensity of 1 W/cm$^2$.  Given these values, we can find the optically modified scattering length as a function of laser intensity and detuning.  Since the bare s-wave scattering length of ${}^{171}$Yb, $a_{bg}=-0.15$a$_0$, is essentially zero, an optical Feshbach resonance will increase $|a|$ by orders of magnitude.

\begin{center}
\begin{table}[b]
\begin{tabular}{|c|c|c|c|}
 \hline
 $E^{theory}_b\, $(MHz) & $E^{exp}_b \, $(MHz) & $\Gamma_M$(kHz) & $l_{opt}$(a$_0$/W/cm$^2$)\\ \hline
3.1  & --- & 243  &  18944 \\	\hline
3.5  & --- & 243  & 804742 \\	\hline
3.6*  & --- & 102  & 641707 \\	\hline
4.6  & --- & 243  & 412804 \\	\hline
7.6  & --- & 243  & 159635 \\	\hline
9.7*  & --- &  97  &  53984 \\	\hline
14.1 & --- & 244  & 63389 \\	\hline
26.8 & 24.7 & 245 & 26887\\   	\hline
34.9* & --- & 89 & 7382\\	\hline
49.9 & 47.8 & 246 & 12571\\  	\hline
89.1  & 86.9 & 248 & 5953\\ 	\hline
152.3 & 149.9 & 251 & 3198\\ 	\hline
250.2 & 247.0 & 256 & 1495\\ 	\hline
396.5 & 391.7 & 261 & 840\\ 	\hline
462.1 & --- & 161 & 13\\ 	\hline
597.3 & --- & 101 & 235 \\ 	\hline
612.5 & 604.3 & 241 & 390 \\ \hline
\end{tabular}
\caption{S-wave accessible excited molecular bound states and resonance properties. $E^{theory}_b$ are the binding energies found from the  multi-channel calculation solutions of the Schr\"{o}dinger equation with Hamiltonian Eq. (\ref{Hamiltonian}). PLR states are marked with *. Binding energies are denoted in frequency units relative to the atomic ${}^1S_0\rightarrow{}^3P_1 (f=3/2)$ transition.  For comparison, $E^{exp}_b$ are the photoassociation resonances observed by Enomoto \emph{et al.} with an experimental uncertainty of $\pm 2$MHz.  The agreement between experiment and theory is good, with a small systematic shift that may be attributed to inaccuracies in the Hund's case-(c) potentials, and/or to systematic shifts in the experimental conditions (e.g. light-shift from the confining dipole potential).  $\Gamma_M$ is the molecular natural linewidth and $l_{opt}$ is the optical length at an intensity of 1 W/cm$^2$.}
\label{tab:swaves} 
\end{table}
\end{center}

\begin{center}
\begin{table}
\begin{tabular}{|c|c|}
 \hline
$E^{theo}_b\, $(MHz) & $E^{exp}_b \, $(MHz) \\ \hline
212.0* & 212.4 \\	\hline
233.8* & 234.0  \\	\hline
258.2 & 256.9  \\	\hline
270.4 & 268.3  \\	\hline
278.9 & 276.8  \\	\hline
355.3* & 355.4  \\	\hline
383.2* & 383.4  \\	\hline
415.5 & 416.1  \\	\hline

\end{tabular}
\hspace{1em}
\begin{tabular}{|c|c|}
\hline
$E^{theo}_b\, $(MHz) & $E^{exp}_b \, $(MHz) \\ \hline
431.9 & 432.0  \\	\hline
443.2 & 442.5  \\	\hline
646.3 & 646.2  \\	\hline
667.7 & 667.2  \\	\hline
682.4 & 681.8  \\	\hline
976.0 & 976.2  \\	\hline
1003.5 & 1002.1  \\	\hline
1022.1 & 1021.4  \\	\hline
\end{tabular}
\caption{P-wave accessible excited molecular bound states. $E^{theo}_b$ are the binding energies found from the  multi-channel calculation as in Table \ref{tab:swaves}. Only the lines which were also experimentally observed by Enomoto \emph{et al.} at a temperature of 25$\mu$K are shown. The experimental uncertainty is $\pm 2$MHz and $\pm 1$ MHz for the PLR states denoted by *.}
\label{tab:pwaves} 
\end{table}
\end{center}

The utility of the OFR for coherent control of nuclear spin exchange depends  on low loss and decoherence.  As a figure of merit, we consider the example of implementing a $\sqrt{SWAP}$ two-qubit entangling unitary between the spin-1/2 nuclei trapped in one site of an optical lattice, as in \cite{anderlini:2007}.  Such a gate occurs if the relative phase shift for singlet vs. triplet nuclear spin states is $\pi/2$. In the perturbative regime and neglecting the small background scattering length, the collisional s-wave phase shift after interaction for time $T$ is $\phi=(4\pi\hbar na_{opt}/\mu)T$, where $n$ is the atomic density.  The loss/decoherence rate is given by $2Kn=8\pi\hbar nb_{opt}/\mu$.  The fidelity of the gate based only on this loss is then
\begin{equation}
\label{fidelity}
 F=e^{-2KnT}=e^{-\left(\frac{\pi}{2}\right)\left(\frac{\Gamma_M}{\Delta}\right)}.
\end{equation}
At the collision energies we consider, p-wave phase shifts are negligible, even in the presence of laser excitation.

To get an estimate of how well the OFR induces nuclear spin exchange, we must balance a variety of constraints.  High fidelity at moderate intensities requires large detuning from molecular resonance, e.g., a fidelity of ~95\% is achieved when $\Delta\approx30\Gamma_M$. Our model, however, assumes sufficiently small detuning so that a single molecular excited state contributes to the resonance.  In addition, to ensure a reasonably fast interaction, the applied intensity must be sufficiently large so that the optical scattering length is large, which will power-broaden the atomic resonance $\Gamma_A\rightarrow\sqrt{1+I/I_{sat}}\Gamma_A$.  For all of these reasons, we consider as an example the photoassociation resonance bound by 396.5 MHz, with a natural linewidth of $\Gamma_M=261$ kHz. This bound state is still relatively close to dissociation, and thus the molecule is long-range, with an outer turning point at 130 a$_0$.  Nonetheless, the density of states is sufficiently sparse that one can detune many linewidths from that resonance while still neglecting coupling to the next higher molecular state, which is bound by 250.2 MHz and has about twice the Franck-Condon factor.  A detuning from the molecular resonance of $\Delta=-30\Gamma_M\approx-7.8$MHz fits this constraint. The intensity is chosen to broaden the resonance so as to increase the scattering length by fixing $\Gamma_{stim}=2k_r l_{opt}\Gamma_M=-\Delta$, or $l_{opt} \approx 10^4$ a$_0$.  From Table \ref{tab:swaves}, this is achieved at an intensity of $I=12.7$ W/cm$^2$, whereby the atomic linewidth is power broadened to $\Gamma_A\rightarrow57$ MHz, which is still narrow compared to a detuning of 400 MHz from dissociation.  When $-\Delta=\Gamma_{stim}\gg\Gamma_{M}$, the optical scattering length is $a_{opt}=-281$ a$_0$ and the loss coefficient is $K \approx 1.15 \times 10^{-12}$ cm$^3$/s.  With this large magnitude scattering length and low loss, given two ${}^{171}$Yb atoms in a lattice site analogous to the experiments at NIST \cite{anderlini:2007} with a density $n=2.4 \times 10^4 \text{cm}^{-3}$, the time of the $\sqrt{SWAP}$ gate is $T\approx 50\mu$s and the gate fidelity is $\approx95\%$.  

In principle, higher fidelity, larger scattering lengths, and shorter gate times are possible by using a higher intensity and larger detuning, though a proper treatment will require the modelling of excitation to multiple molecular bound states and line broadening.  Even with these modest parameters, we see that OFRs have great potential for control of nuclear-spin exchange and strong entangling interactions.  The combination of this tool with recent advances in loading optical lattices via superfluid to mott insulator phase transition \cite{fukuhara:2009}, the ability to optically manipulate nuclear spin coherence \cite{boyd:2006} and re-cool atoms without decohering nuclear spins \cite{reichenbach:2007}, and proposals for quantum logic \cite{hayes:2007, daley:2009} make this system attractive for new applications in quantum information processing.

{\bf Acknowledgements}
We want to thank Eita Tiesinga for valuable discussions.
This work was supported by the Office of Naval Research, Grant No. N00014-03-1-0508, and IARPA Grant No. DAAD19-13-R-0011.


\begin{thebibliography}{50}
  \bibitem{boyd:2006} M. M. Boyd {\em et al.}, Science {\bf 314} 1430 (2006).

  \bibitem{hayes:2007} D. Hayes, P. S. Julienne, and I. H. Deutsch, Phys. Rev. Lett. {\bf 98}, 070501 (2007).
  
 \bibitem{reichenbach:2007} I. Reichenbach and I. H. Deutsch, Phys. Rev. Lett. {\bf 99} 123001, (2007).
 
 \bibitem{daley:2009} A. J. Daley  {\em et al.}, Phys. Rev. Lett. {\bf 102}, 040402 (2009).  A. V. Gorshkov{\em et al.},  {\em ibid}, 110503 (2009).
 
 \bibitem{gorshkov:2009} A. V. Gorshkov {\em et al.}, e-print arXiv:0905.2610. 

  \bibitem{chin:2009} C. Chin {\em et al.}, e-print arXiv:0812.1496.

  \bibitem{bloch:2008} I. Bloch, J. Dalibard, and W. Zwerger, Rev. Mod. Phys. {\bf 80}, 885 (2008).

  \bibitem{giorgini:2008} S. Giorgini, L. Pitaevskii, and S. Stringari, Rev. Mod. Phys. {\bf 80}, 1215 (2008)
   
  \bibitem{bourdel:2004} T. Bourdel  {\em et al.}, Phys. Rev. Lett. {\bf 93}, 050401 (2004).
  
  \bibitem{regal:2003} C. A. Regal {\em et. al}, Nature {\bf 424}, 47 (2003).
  
   \bibitem{bauer:2009} D. M. Bauer {\em et. al}, Nature Phys. {\bf 5}, 339 (2009).

   \bibitem{bohn:1997} J. L. Bohn and P. S. Julienne, Phys. Rev. A {\bf 56}, 1486 (1997).

  \bibitem{ciurylo:2005} R. Ciury\l{}o, E. Tiesinga, and P. S. Julienne, Phys. Rev. A {\bf 71}, 030701 (2005).
  
  \bibitem{jones:2006} K. M. Jones  {\em et al.}, Rev. Mod. Phys. {\bf 78} 483 (2006). 

  \bibitem{fatemi:2000} F. K. Fatemi, K. M. Jones and  P. D. Lett, Phys. Rev. Lett. {\bf 85}, 4462 (2000).

  \bibitem{theis:2004} M. Theis  {\em et al.}, Phys. Rev. Lett. {\bf 93}, 123001 (2004).

  \bibitem{thalhammer:2005} G. Thalhammer  {\em et al.}, Phys. Rev. A {\bf 71}, 033403 (2005).
  
 \bibitem{enomoto:2008} K. Enomoto  {\em et al.}, Phys. Rev. Lett. {\bf 100} (2008).

  \bibitem{zelevinsky:2006} T. Zelevinsky  {\em et al.}, Phys. Rev. Lett. {\bf 96}, 203201 (2006).
  
\bibitem{enomoto:2008_2} K. Enomoto  {\em et al.}, Phys. Rev. Lett. {\bf 101}, 203201 (2008).
  
   \bibitem{killian:2009} Y. N. Martinez de Escobary  {\em et al.}, (unpublished).
 
   \bibitem{anderlini:2007} M. Anderlini  {\em et al.}, Nature {\bf 448}, 452 (2007).

  \bibitem{gao:1998} B. Gao, Phys. Rev. A {\bf 58}, 1728 (1998).

  \bibitem{kitagawa:2008} M. Kitagawa  {\em et al.}, Phys. Rev. A {\bf 77}, 012719 (2008).
 
  \bibitem{tiesinga:2005} E. Tiesinga  {\em et al.}, Phys. Rev. A {\bf 71} (2005).
  
  \bibitem{colbert:1982} D. T. Colbert and W. H. Miller, J. Chem. Phys. {\bf 96}, 1982 (1992).
  
  \bibitem{fukuhara:2009} T. Fukuhara  {\em et al.}, Phys. Rev. A {\bf 79}, 041604R (2009).

 
 
 
\end{thebibliography}
\end{document}